\documentclass[]{aa}
\usepackage{graphicx, natbib}
\usepackage{txfonts}
\bibpunct{(}{)}{;}{a}{}{,}
\def \kms {$\mathrm{km\;s^{-1}\;}$}
\begin{document}
\title{Spectropolarimetry of SN 2006aj at 9.6 days\thanks{Based on
observations made with ESO Telescopes at the Paranal Observatory,
under programme 76.D-0177(A).}}  
\author{Justyn R. Maund \inst{1} \and
J. Craig Wheeler \inst{1} \and Ferdinando Patat \inst{2} \and Dietrich
Baade \inst{2} \and Lifan Wang \inst{3} \and Peter H\"{o}flich \inst{4}}
\offprints{Justyn Maund, \email{jrm@astro.as.utexas.edu}}

\institute{Dept. of Astronomy and McDonald Observatory, The University
of Texas at Austin, 1 University Station, C1400, Austin, Texas
78712-0259, U.S.A.
\and ESO - European Organisation for Astronomical Research in the
Southern Hemisphere, Karl-Schwarzschild-Str.\ 2, 85748 Garching b.\
M\"unchen, Germany
\and Department of Physics, Texas A\&M University, College Station,
Texas 77843-4242, U.S.A.
\and Department of Physics, Florida State University, Tallahassee,
Florida 32306-4350, U.S.A.\ } 
 \date{Received /Accepted }

\abstract{The observational technique of spectropolarimetry has been
used to directly measure the asymmetries of Supernovae (SNe),
Gamma-Ray Bursts (GRBs) and X-Ray Flashes (XRFs).}  {We wish to
determine if non-axial asymmetries are present in SNe that are
associated with GRBs and XRFs, given the particular alignment of the
jet axis and axis of symmetry with the line of sight in these cases.}
{We performed spectropolarimetry with the Very Large Telescope (VLT)
FORS1 instrument of the Type Ic SN 2006aj, associated with the XRF
060218, at V-band maximum at 9.6 rest frame days after the detection
of the XRF.  Due to observations at only 3 retarder plate angles, the
data were reduced assuming that the instrumental signature
correction for the $U$ Stokes parameter was identical to the
correction measured for $Q$.}  {We find SN 2006aj to be highly
polarized at wavelengths corresponding to the absorption minima of
certain spectral lines, particularly strong for O I 7774\AA\ and Fe
II, observed at 4200\AA\ with a polarization 3\%.  The value of the
Interstellar Polarization is not well constrained by these
observations and, considering the low polarization observed between
6000-6500\AA, the global asymmetry of the SN is $\lesssim 15\%$.  O I
and Fe II lines share a polarization angle that differs from Ca II.}
{SN 2006aj exhibits a higher degree of line polarization than other
SNe associated with GRBs and XRFs.  The polarization associated with
spectral lines implies significant asymmetries of these elements with
respect to each other and to the line of sight.  This is contrary to
the standard picture of SNe associated with GRBs/XRFs, where the axis
of symmetry of the SN is aligned with the GRB jet axis and the line of
sight.}
\maketitle
\section{Introduction}
\label{section:introduction}
The number of observed core-collapse Supernovae (CCSNe) associated
with Gamma Ray Bursts (GRBs) and X-ray Flashes (XRFs) is still low and
there are few examples of such SNe which have also been observed at
early times with spectropolarimetry: SN 1998bw (GRB 980425;
\citealt{2001ApJ...555..900P}) and SN 2003dh (GRB 030329;
\citealt{2003ApJ...593L..19K}).  Here we present spectropolarimetry
of the Type Ic SN 2006aj associated with XRF 060218.\\ XRF 060218 was
detected on 2006 Feb 18.15 UT by the Swift Burst Alert Telescope
\citep{2006GCN..4775....1C}, and located at
$\mathrm{\alpha_{J2000}=03^{h}21^{m}39.7^{s}}$ and
$\mathrm{\delta_{J2000}=+16\degr52\arcmin01.8\arcsec}$ by the Swift
X-ray Telescope \citep{2006GCN..4786....1C}.  The XRF had a long
duration ($\sim3000$s; \citealt{2007MNRAS.375L..36G}), and was
reported as being under-luminous and softer than normal GRBs
\citep{2006Natur.442.1011P}.  \citet{2006GCN..4792....1M} reported
optical spectroscopy of the afterglow, at Feb 20.04, revealing nebular
emission features from the host galaxy with $z=0.033$.
\citet{2006GCN..4809....1F} commented that optical spectroscopy of the
afterglow revealed the presence of a Type Ic SN similar to SN 1997ef,
and this SN was subsequently labelled SN 2006aj
\citep{2006IAUC.8674....2S}.  \citet{2006Natur.442.1011P} and
\citet{2006ApJ...645L..21M} observed SN 2006aj to have a rapid rise to
V-band maximum in 10 days.  \citet{2006A&A...459L..33G} observed
evolution in the R-band linear polarization, from a rapidly evolving
high polarization ($\sim4.5\%$ at 3 days following the GRB) and a
lower polarization ($\sim 1.4\%$) after the V-band maximum of the SN
light curve.  Late-time spectropolarimetry of SN 2006aj (at 206 rest
frame days after the XRF) showed no polarization to $2\%$ at
$3\sigma$, while observations of the nebular spectrum revealed
asymmetric line profiles for $\mathrm{[O I]}\lambda\lambda 6300,6363$
\citep{2007ApJ...661..892M}.
\section{Observations}
\label{section:observation}
Spectropolarimetry of SN 2006aj was acquired on 2006 Feb 28.02, using
the European Southern Observatory (ESO) Very Large Telescope (VLT)
FOcal Reducer and low dispersion Spectrograph (FORS1) instrument in
the spectropolarimetry ``PMOS'' mode \citep{1998Msngr..94....1A}.
This corresponds to 9.87 days following original detection of the XRF,
and 9.55 days in the rest frame.  At this epoch, SN 2006aj was
approaching the solar position, limiting the available observing
time. Due to this constraint, a full spectropolarimetry dataset was
not acquired and, due to the nature of spectropolarimetry
observations, this dataset could not be completed at a later epoch.
Observations of SN 2006aj were acquired with the retarder plate at
only three positions: 0\fdg0, 45\fdg0 and 22\fdg5.  Exposure times
were 900s, with the retarder plate positioned at 0\fdg0 and 45\fdg0,
and 793s, at 22\fdg5.  The G300V grism was used, with pixel scale
$\mathrm{2.6\AA\;px^{-1}}$ and spectral resolution, measured from arc
lamp calibration spectra, of 12.6\AA.  The data were reduced in the
standard manner, following the routine outlined by \citet{maund05bf}.
Due to incomplete observations, the Stokes parameters were determined
following the technique of \citet{myresnote}, which assumes that the
``instrumental signature corrections'' of the $Q$ and $U$ Stokes
parameters, $\epsilon_{Q}$ and $\epsilon_{U}$, are approximately the
same.\\ Given the broadness of the spectroscopic features, due to high
expansion velocities ($\gtrsim \mathrm{\sim 18\;000\, km\, s^{-1}}$;
\citealt{2006Natur.442.1011P}), the Stokes parameters were coarsely
re-sampled to 50\AA\ wavelength bins.  A correction for the
recessional velocity of the host galaxy was applied to the data, using
$z=0.033$ derived from the narrow nebular lines observed in this data
(this is completely consistent with previously reported measures of
the recessional velocity; e.g. \citealt{2006GCN..4792....1M}).
\section{Results and Analysis}
\label{section:results}
\begin{figure*}
\centering
\includegraphics[width=12cm, angle=270]{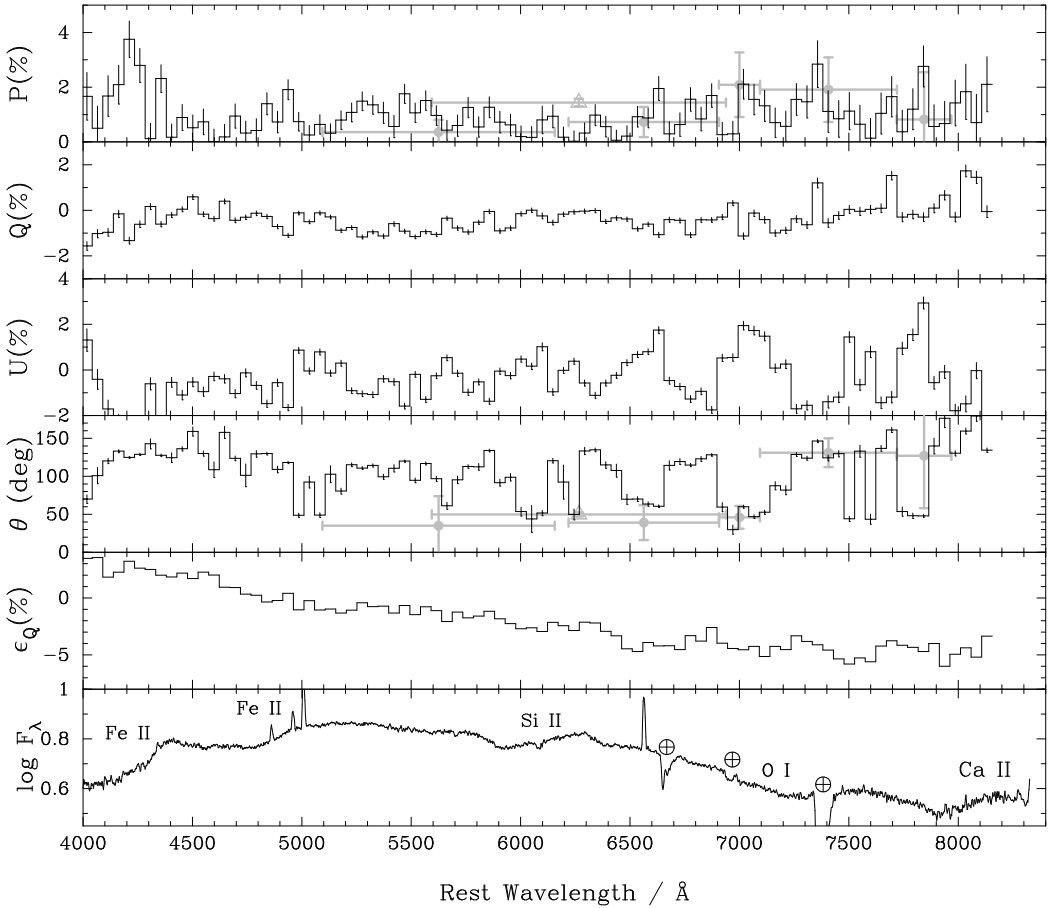}
\caption{Spectropolarimetry of SN 2006aj at 9.55 days. {\it From top
    to bottom:} total polarization $p$, the $Q$ Stokes parameters, the
    $U$ Stokes parameter, the polarization angle $\theta$, the
    instrumental signature correction $\epsilon_{Q}$ observed for
    the $Q$ Stokes parameters and the flux spectrum ($\mathrm{ergs\;
    s^{-1} \; cm^{-2} \AA^{-1}}$).  The wavelength scales for the
    polarization parameters have been rebinned to 50\AA\ bin$^{-1}$,
    and the flux spectrum is binned at 2.6\AA\ bin$^{-1}$.  Also shown
    are polarization parameters of
    \citet{2006A&A...459L..33G}($\bigtriangleup$) for SN 2006aj at
    13.7 days and \citet{2003ApJ...593L..19K}($\bullet$) for SN 2003dh
    at 16 days after V-band maximum.  The wavelength scales for each
    of the data sets have been corrected for the recessional
    velocities of the respective host galaxies.  The narrow lines in
    the flux spectrum ({\it bottom panel}) are nebular lines arising
    in the host galaxy.  Telluric absorption features are indicated by
    $\oplus$.}
\label{fig:results:panel}
\end{figure*}

\begin{figure}
\centering
\includegraphics[width=8cm,angle=270]{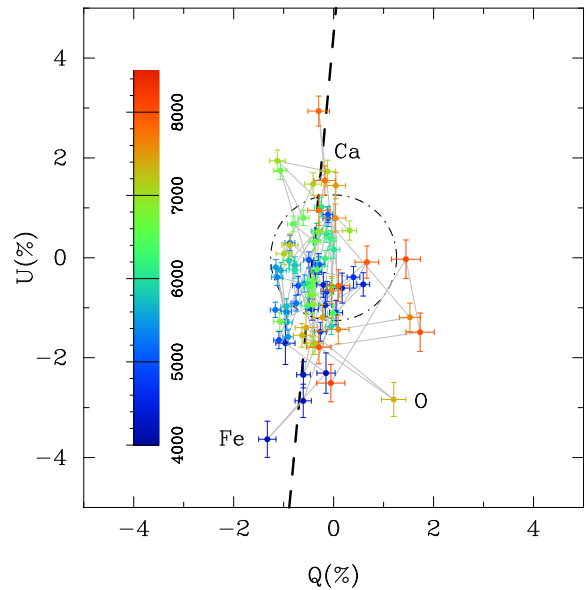}
\caption{The observed Stokes parameters of SN 2006aj, on the $Q-U$ plane.  
The data have been rebinned to 50\AA, but have not been corrected for an
ISP component.  The maximum ISP $p<1.26\%$, corresponding to the
foreground Galactic extinction of $E(B-V)=0.14$, is indicated by the
dot-dashed circle.  The points are colour-coded by wavelength, according
to the scheme of the colour bar.  The heavy dashed line indicates the
dominant axis of the data \citep{2001ApJ...550.1030W}.}
\label{fig:results:qupanel}
\end{figure}

The observed Stokes parameters and flux spectrum of SN 2006aj at 9.55 
days are presented as Fig. \ref{fig:results:panel}.  This observation 
was acquired at the maximum of the V-band ($m_{V}=17.5$; \citealt{2006ApJ...645L..21M}) and bolometric light curves 
of the SN \citep{2006A&A...459L..33G,2006Natur.442.1011P}.
The flux spectrum is composed of a series of broad overlapping
absorption and emission features; the flux spectrum has been
extensively discussed by \citet{2006Natur.442.1011P} and
\citet{2006ApJ...645L..21M}.  The polarization spectrum is observed to
be variable over the wavelength range of 1 bin.  The wavelength range
of 6000-6500\AA\ is typified by a low level of polarization of
$0.5\pm0.3\%$ (null polarization at a $2\sigma$ level).  Over the
larger range of 4500-6500\AA, the polarization is $\lesssim 2\%$.  The
highest degrees of polarization are 2-4\% and associated with features
at the blue and the red extreme of the spectrum.  A
wavelength-dependent Interstellar Polarization (ISP) component may be
responsible for some of this polarization, and is discussed later, but
it is clear that significant variability of the polarization spectrum
over small wavelength ranges implies that there is significant
polarization intrinsic to SN 2006aj itself.  In addition, we note that
the bulk of the polarization signature derives from the $U$ Stokes
parameter, but the features are not correlated with the
$Q-$instrumental signature correction.\\ The R-band polarimetry of
\citet{2006A&A...459L..33G}, at 13 days after the XRF, is $2\sigma$
higher than the spectropolarimetry of three days prior (see
Fig. \ref{fig:results:panel}).  There is some variability in the
degree of polarization across the observed wavelength range of the
R-band, but there appears no obvious wavelength dependence that would
contribute to the broad-band measurement of
\citeauthor{2006A&A...459L..33G}.  A region of higher polarization is
observed immediately redward of 7000\AA\ in the rest frame and would
have contributed to the R-band polarization if the recessional
velocity of the host galaxy were lower.  Differences between the
polarization and the polarization angle of the spectropolarimetry and
the observations of \citet{2006A&A...459L..33G} are most likely due to
the differences in the epoch at which the respective observations were
conducted.\\ The rise in the polarization at the red extreme of the
spectrum is most likely associated with O I $\lambda7774$\ and the Ca
II IR triplet.  There are particular peaks in the polarization
spectrum which, given the correspondence of polarization peaks and
absorption minima of P Cygni profiles in SN spectra
\citep{1984MNRAS.210..829M}, can be associated with the absorption
minima of these lines.  The polarization peaks correspond to
absorption velocities of O $\lambda7774$\ at
$29\;200\;\pm\;2\;000$\kms and Ca II IR triplet
$25\;800\;\pm\;1\;800$\kms.  The large uncertainties on these
velocities arise from the coarse rebinning of the Stokes parameters to
50\AA.  At the blue end of the spectrum, the polarization of
$4\pm0.7\%$ at 4200\AA\ is associated with Fe II multiplets 37 and 38
at $20\;500\pm 3\;500$\kms.  We suggest that peaks in the polarization
between 4500-5000\AA\ are due to redder Fe II lines, particularly that
of Fe II $\lambda5169$\ at 4850\AA, implying a velocity of
$18\;500\;\pm 3000$\kms.  The lower polarization of the redder Fe II
lines is not unexpected, as the bluer Fe II lines of multiplets 37 and
38 have been observed to produce a single P Cygni profile, given the
small difference in wavelength of the constituent lines and the high
velocities observed in SNe, whereas the redder lines can be resolved
\citep{maund05bf}.  We note that these velocities are similar to the
values quoted by \citet{2006Natur.442.1011P} for the photospheric
velocity, measured with the Fe lines, and \citet{2006Natur.442.1018M},
who consider an O-dominated shell of ejecta at $20\;000 \lesssim
{\mathrm v} \lesssim 30\;000$\kms.  There are two minor peaks in
polarization between $\mathrm{5500\AA < \lambda <6000\AA}$ which may
be associated with Si II $\lambda6355$, that
\citet{2006Natur.442.1018M} identified in spectra.\\ There is an
additional apparent peak in the polarization spectrum at 7840\AA.  The
measured polarization, at this wavelength, has a high uncertainty, due
to being coincident with a low S/N portion of the spectrum at a broad
telluric absorption.  This feature is not, therefore, intrinsic to the
SN, nor significantly more polarized than the immediately surrounding
wavelengths.\\ The observed Stokes parameters are plotted on the $Q-U$
plane on Fig.  \ref{fig:results:qupanel}.  The data conform to an
obvious dominant axis, which is elongated along the $U$ axis.  Some
degree of elongation along the $U$ axis is to be expected, since the
errors on the $U$ Stokes parameter are generally double those
determined for $Q$.  This elongation is, however, substantially larger
than the uncertainties on the $U$ parameter, suggesting it is real and
not just due to significant differences between $\epsilon_{Q}$ and
$\epsilon_{U}$.  No obvious loops are observed, corresponding to the
Stokes parameters across particular spectral lines, on the $Q-U$ plane
\citep{maund05bf}.  This is most likely due to the coarse rebinning
and relatively low level of S/N ($\sim 200$).  The Stokes parameters
at wavelengths corresponding to different spectral lines are, however,
observed to occupy different portions of the $Q-U$ plane.  The
polarization angles of the Fe II and O I features are similar and
these spectral features occupy approximately the same location ($U<0$)
on the $Q-U$ plane (see Fig. \ref{fig:results:qupanel}).  The Ca II IR
triplet is distinct from the Fe II and O I features, with a
polarization angle offset from Fe II and O I by $\sim90\degr$ and
occupying the $U>0$ region of the $Q-U$ plane.  The dominant axis is
principally defined by those features with the highest polarizations,
which correspond to the Fe II, O I and Ca II absorption features.  The
dominant axis is offset from the origin of the $Q-U$ plane, which can
be due to the ISP or can be intrinsic to the SN itself
\citep{my2001ig}.\\
\citet{schleg98}\footnote{http://nedwww.ipac.caltech.edu/} give the
foreground Galactic reddening of $E(B-V)=0.14$, which corresponds to
$p_{ISP}<1.26\%$.  \citet{2007ApJ...661..892M} report late-time
spectropolarimetry of SN 2006aj, with an upper limit on the total
polarization of $p(3\sigma) <2\%$ (at late times, as the density of
scattering particles decreases $p\rightarrow p_{ISP}$), while
\citet{2006A&A...459L..33G} suggest that the observed R-band
polarization of $1.4\%$ at days $> 13$ is consistent with
$E(B-V)>0.15$.  We note that the data in the wavelength range of
6000-6500\AA\ suggests $E(B-V)>0.05$, for no intrinsic polarization or
ISP depolarization.\\ There is a peak in the polarization at 5500\AA,
which is not matched by an obvious absorption feature in the flux
spectrum.  This may well represent the maximum of the ISP, where
$p=1.4\pm0.4\%$, which would be consistent with the ISP estimated by
\citeauthor{2006A&A...459L..33G}.
\section{Discussion and Conclusions}
\label{section:discussion}
  It is particularly important to determine whether the assumption
that $\epsilon_{U}\approx\epsilon_{Q}$ is correct.  It is clear from
comparison with the data of \citet{2006A&A...459L..33G} that this
assumption is indeed approximately correct, to within $\sim 1\%$.
Examination of the three exposures, at each of the retarder plate
positions, suggests that the observations were not in the regimes
identified by \citet{myresnote} for which $\epsilon_{Q}$ and
$\epsilon_{U}$ are expected to diverge: large gain gradient between
the ordinary and extraordinary rays and polarizations $\gtrsim 20\%$.
Monte Carlo models of the observations, given a signal-to-noise ratio
of $\sim200$ at each retarder plate angle, yielded an error on the $U$
instrumental signature correction $\Delta \epsilon_{U}\sim\pm0.3\%$.
This approach is limited, however, as any defects peculiar to the
observation with the retarder plate at 22\fdg5 cannot be identified,
having not obtained an observation at 67\fdg5.  As the Stokes
parameters are calculated using normalised flux differences, the
corrections are also flux normalized; the difference in exposures
times only leads to an increase in the uncertainty on $U$ due to
Poisson noise as well as $\Delta (\epsilon_{Q}-\epsilon_{U})$.\\ In
addition, in the ESO archive there are seven spectropolarimetry
datasets of SNe in the 19 days prior to XRF 060218/SN 2006aj and two
sets in the month following \citep{nando2006X} which, with complete
observations at all four retarder plate positions, give errors on the
$U$ instrumental signature corrections, $\epsilon_{U}$, consistent
with the signal-to-noise and the model presented by \citet{myresnote}.
These observations show no obvious time dependent defects over the
time period immediately surrounding this observation of SN 2006aj.
The caveat that the Stokes parameters were not determined completely
independently must always be remembered when considering these
results.\\ The wavelength region 5500-6500\AA\ shows the
smallest degree of polarization variability in the polarization
spectrum, where the $p\sim0.45\%$.  A similar polarization spectrum
was observed for SN 1998bw/GRB 980425, with little wavelength
dependence across the same wavelength range (for observations at 7
days before and 10 days after V maximum, although the level of
polarization was lower at the later epoch;
\citealt{2001ApJ...555..900P}).  The polarization across this
wavelength range can be used to place an upper limit on the degree of
global asymmetry of SN 2006aj, assuming that all the polarization is
intrinsic to the SN. Comparison with \citet{1991A&A...246..481H}
shows that this level of polarization requires deviations from a
spherical symmetry of $\lesssim 15\%$.  In the presence of a
substantial ISP, such as that discussed by
\citet{2006A&A...459L..33G}, the degree of the asymmetry of the
photosphere would tend to zero.  Alternatively, if the polarization
angle of the ISP is fortuitously counter aligned with the polarization
angle of the intrinsic SN polarization, then the ISP would effectively
depolarize the SN.  This would yield an underestimation for the
asymmetry of SN 2006aj.  Regardless of the ISP, however, the
wavelength-dependence of the polarization, especially coincident with
absorption features in the flux spectrum, is not consistent with a
smoothly varying ISP law.  This demonstrates that asymmetries are
significantly present in SN 2006aj.\\ The high degree of polarization
associated with O I $\lambda7774$\ and the Ca II IR triplet in the
data of SN 2006aj is similar to that observed in high-velocity SNe and
other SNe associated with GRBs, such as SN 2003dh/GRB 030329.  For SN
2003dh there is a clear rise in polarization at the red end of the
spectrum associated with the approximate location of the O I $\lambda
7774$\ line, as shown on Fig. \ref{fig:results:panel}.  The data of
\citet{2003ApJ...593L..19K} was, however, acquired at $\sim16$ days
after the corresponding V-maximum (18 days after the GRB 030329;
\citealt{2003ApJ...599..394M}).  The wavelength range containing the
Fe II lines was not covered with the spectropolarimetry observations
of SN 1998bw and SN 2003dh.  The Fe II lines were observed in SN
2002ap and were less polarized, at $0.3\%$ at V-maximum, than those
observed in SN 2006aj, and substantially less polarized than the
corresponding O I 7774\AA\ line at 1\% \citep{2003ApJ...592..457W}.
The Ca II IR triplet in SN 2006aj is also more polarized than observed
in SN 2003dh \citep{2003ApJ...593L..19K}.  \\ SN 2006aj does not show
conspicuously higher polarization than other examples of CCSNe.  Some
``stripped-core'' CCSNe show higher levels of polarization of their
core layers, implying higher degree of asymmetries, over larger
wavelength ranges than observed for SN 2006aj \citep{my2001ig}.  The
highest polarization observed for a Type Ic SN was $\gtrsim 4\%$ for
SN 1997X \citep{2001ApJ...550.1030W}.  In the cases of future, better
observed CCSNe associated with GRBs, for which the ISP can be more
thoroughly measured, the intrinsic polarizations and polarization
angles may have similarities given the preferred orientation of such
objects.\\ The polarization angles of the Fe II and Ca II lines are
significantly different, and subtraction of an ISP, with constant
polarization angle, would change the absolute values of the
polarization angles of these features, but would not eliminate their
relative offset.  The dissimilar polarization angles is suggestive of
different geometries of these species within the ejecta.
\citet{2007ApJ...661..892M} note that spectral synthesis models of the
SN 2006aj suggest that the O I, in the late time ejecta, is configured
in a face-on disk, while Ni has an almost spherical geometry.  Neither
of these geometries provides the distinct non-axisymmetric component
which would yield such high levels of polarization.\\ The presence of
strong polarization associated with lines points to an asymmetric
distribution of these elements with respect to the angle at which SN
2006aj is observed.  Importantly, SN 2006aj was observed with a
specific orientation, with the line of sight and the jet axis being
coincident.  \citet{2006Natur.442.1011P} rule out the scenario of XRF
060218 having been observed significantly off-axis, such that an
axisymmetric SN would appear polarized due to inclination effects.  If
the jet axis defines the axial symmetry, then the presence of
asymmetry as observed as polarization is difficult to rectify with the
axisymmetric standard model of the SNe-GRBs \citep{2006ARA&A..44..507W}. 
 Alternatively, if the jet opening angle is large (e.g.  
$\theta_{j}>80\degr$ measured by \citet{2006Natur.442.1014S} from the
radio light curve of SN 2006aj) the line of sight may be significanty
inclined with respect to the expected axis of symmetry of the progenitor
and the SN, but still allow the prompt emission to be observed.\\
The wavelength-dependent polarization observed, at later epochs, for SN
2003dh implies that the presence of such asymmetries is not limited to SN
2006aj, and is a property of SNe associated with GRBs.


\bibliographystyle{aa}

\end{document}